\documentclass[12pt,preprint]{aastex}

\shorttitle{A Redshift for \object {ESO 243-49 HLX-1}}
\shortauthors{Wiersema et al.}

\begin{document}


\title{A Redshift for the Intermediate Mass Black Hole Candidate HLX-1: Confirmation of its Association with the Galaxy \\ESO 243-49}


\author{K. Wiersema$^1$, S. A. Farrell$^1$, N. A. Webb$^{2,3}$, M. Servillat$^{4}$, T. J. Maccarone$^{5}$, D. Barret$^{2,3}$ and O. Godet$^{2,3}$}
\affil{$^1$ Department of Physics and Astronomy, University of Leicester, University Road, Leicester, LE1 7RH, UK}
\affil{$^2$ Universit\'{e} de Toulouse, UPS, CESR, 9 Avenue du Colonel Roche, F-31028 Toulouse cedex 9, France}
\affil{$^3$ CNRS, UMR5187, F-31028 Toulouse, France}
\affil{$^4$ Harvard-Smithsonian Center for Astrophysics, 60 Garden Street, MS-67, Cambridge, MA 02138, USA}
\affil{$^5$ School of Physics and Astronomy, University of Southampton, Hampshire, SO17 1BJ, UK}



\begin{abstract}
In this Letter we report a spectroscopic confirmation of the association of \object{HLX-1}, the brightest ultra-luminous X-ray source, with the galaxy ESO~243-49. At the host galaxy distance of 95 Mpc, the maximum observed 0.2 -- 10 keV luminosity is 1.2 $\times$ 10$^{42}$ erg s$^{-1}$. This luminosity is $\sim$400 times above the Eddington limit for a 20 M$_\odot$ black hole, and has been interpreted as implying an accreting intermediate mass black hole with a mass in excess of 500 M$_\odot$ (assuming the luminosity is a factor of 10 above the Eddington value). However, a number of other ultra-luminous X-ray sources have been later identified as background active galaxies or foreground sources. It has recently been claimed that \object{HLX-1} could be a quiescent neutron star X-ray binary at a Galactic distance of only 2.5 kpc, so a definitive association with the host galaxy is crucial in order to confirm the nature of the object. Here we report the detection of the H$\alpha$ emission line for the recently identified optical counterpart at a redshift consistent with that of ESO 243-49. This finding definitively places \object{HLX-1} inside ESO 243-49, confirming the extreme maximum luminosity and strengthening the case for it containing an accreting intermediate mass black hole of more than 500 M$_\odot$.\end{abstract}





\section{Introduction}

Ultra-luminous X-ray (ULX) sources are extragalactic objects located outside the nucleus of the host galaxy with bolometric luminosities $>$ 2.6 $\times$ 10$^{39}$ erg s$^{-1}$ \citep[e.g.][]{rob07}. These extreme luminosities appear to exceed the theoretical Eddington limit for a stellar mass black hole, where the gravitational pressure of in-falling material balances against the radiation pressure, implying the presence of accreting black holes with masses between $\sim$10$^2$ -- 10$^5$ M$_\odot$. The existence of such intermediate mass black holes is widely disputed, as hyper-accretion \citep{beg02} and/or beaming \citep{koe02,fre06,kin08} can cause a stellar mass black hole to appear to exceed the Eddington limit. However, the rare sub-class of hyper-ULXs with luminosities $>$ 10$^{41}$ erg s$^{-1}$ require extreme tuning to explain without the presence of an intermediate mass black hole \citep[e.g][]{rob07,mil09}, needing either non-physical levels of super-Eddington accretion or very small beaming factors. The luminosity derived for these objects is entirely dependent upon the X-ray source being located within the host galaxy so ruling out background \citep[e.g.][]{cla05} or foreground \citep[e.g.][]{wei04} sources is essential. This is normally not possible to achieve with the X-ray data alone, and so multi-wavelength follow-up observations are required in order to confirm the ULX nature. For example, the previous title-holder for the brightest ULX in the galaxy MCG-03-34-63 \citep{min06} has recently been found to be a background Active Galactic Nucleus (AGN; Miniutti et al. in preparation).

The brightest ULX candidate currently known is the X-ray source \object{HLX-1} coincident with the galaxy \object{ESO 243-49}, with a derived maximum luminosity (assuming it is at the galaxy redshift of z = 0.0224) of $\sim$10$^{42}$ erg s$^{-1}$ \citep{far09}. This object is almost an order of magnitude more luminous than the other hyper-ULXs \citep[e.g.][]{gao03}, and would provide strong evidence for the existence of intermediate mass black holes if the luminosity can be confirmed. Following its discovery, \object{HLX-1} has been regularly monitored in X-rays by the \emph{Swift} observatory, and has been found to vary by a factor of $\sim$20 over relatively short time scales, at the same time exhibiting spectral variability similar to the canonical state transitions seen in Galactic stellar mass black hole binaries \citep{god09}. The X-ray source position was recently improved to sub-arcsecond accuracy using data from the \emph{Chandra} space telescope \citep{web10}, allowing a faint optical counterpart (R = 23.8 mag) to be identified \citep{sor10}. The maximum X-ray to optical flux ratio (F$_x$/F$_{opt}$) of \object{HLX-1} is $\sim$1000, two orders of magnitude higher than is typically seen for AGN \citep[e.g][]{sev03}, making a background source highly unlikely. However, it has recently been argued that the X-ray spectrum and F$_x$/F$_{opt}$ ratio of \object{HLX-1} were consistent with a Galactic quiescent neutron star X-ray binary with a late-type secondary at a distance of $\sim$2.5 kpc \citep{sor10}. These objects contain neutron stars accreting at low rates from a low-mass companion star, and have high F$_x$/F$_{opt}$ ratios and X-ray spectra that are typically dominated by thermal emission from the surface or atmosphere of the neutron star \citep[e.g.][]{web07,jon08}. 

In this Letter we present spectroscopic observations of \object{HLX-1} with the Very Large Telescope (VLT). In $\S$ 2 we describe the observations and in $\S$3 we outline the data reduction and analysis steps. $\S$ 4 reports the results that we obtained, and $\S$ 5 presents a discussion and the conclusions that we have drawn.  

\section{Observations}

\begin{figure}
\epsscale{1}
\plotone{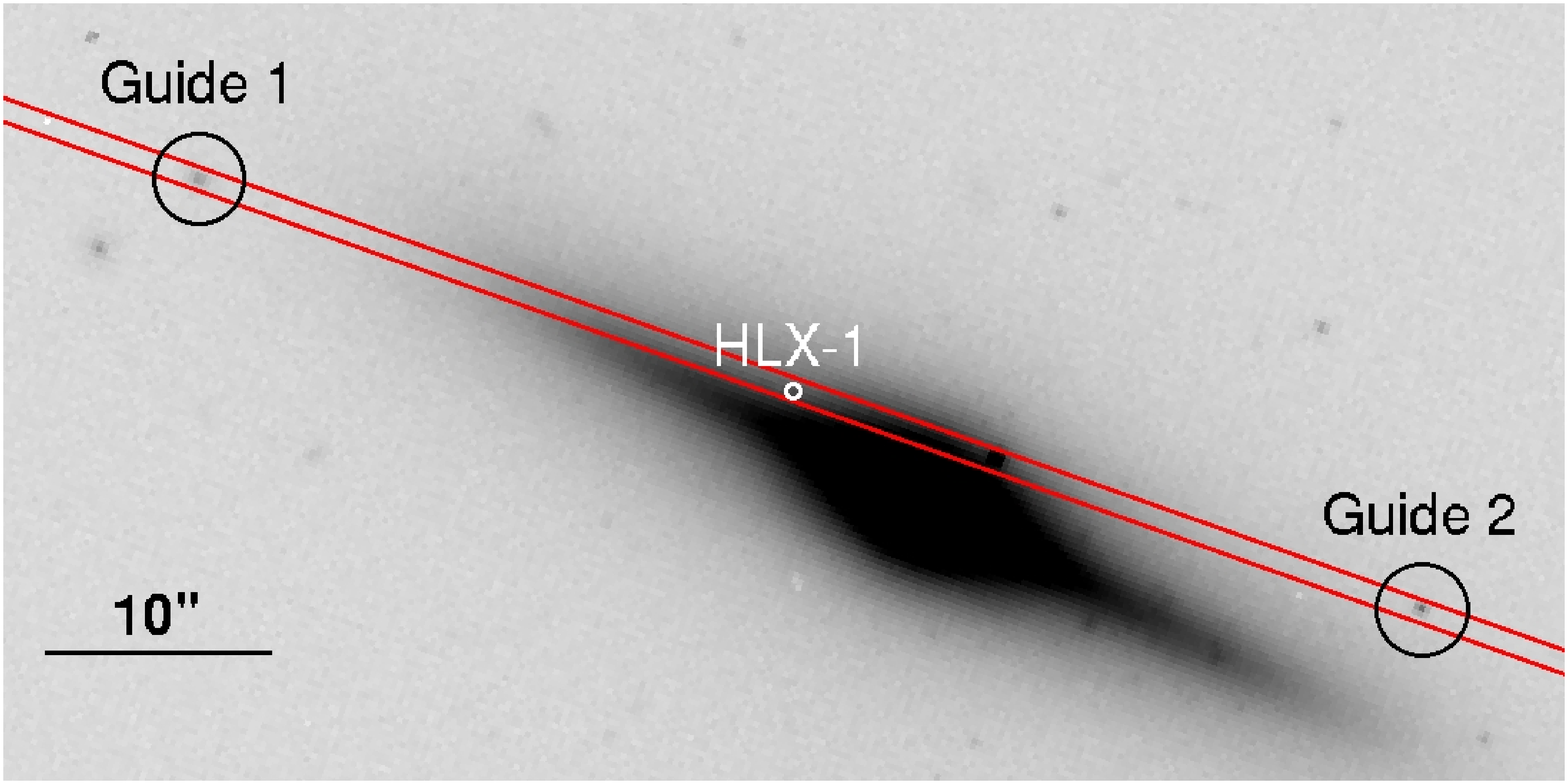}
\caption{I-band VLT optical image of ESO 243-49. The slit position (red box) with a width of 1\arcsec~was aligned on two guide stars that fortuitously lined up with the \emph{Chandra} position of HLX-1 (white circle, with the radius corresponding to the 0.3$\arcsec$ positional error).\label{slit}}
\end{figure}

Following the detection of the counterpart we obtained optical spectroscopic data using the VLT accorded under the Directors Discretionary Time (DDT) program (ID: 284.D-5008). The VLT acquisition absolute astrometry is not good enough to acquire blindly a given position in RA and Dec when the source is not detectable by the acquisition camera, as the typical uncertainties exceed the slit width. A commonly used method is to use blind offsets from the position of a bright reference star. The offset star is acquired accurately using a through-the-slit exposure after which the blind offset is performed. The typical uncertainty associated with this method is $\sim$0.3$\arcsec$ or better. However, the most accurate method is to use fortuitous alignment of foreground stars with the target. With long slits such as that of FORS2, the chances of two random bright objects aligning with the target is fairly large, provided that there are no strict requirements for the  slit position angle or parallactic angle. To find such alignments, we used a deep image of the field in the J band (theoretical detection limit of J = 22.5 mag) that we obtained with PANIC at the Baade Magellan Telescope on 2009 August 8. We found two field stars of J $\sim$ 20 mag that allowed a precise positioning of the long slit of FORS2 so that it would cover the position of HLX-1 (see Figure \ref{slit}). The two stars can be centered in the slit very accurately (to the accuracy of the centroid determination of each star, so considerably better than 1 pixel). 

We obtained deep spectroscopy on four nights, using the FORS2 instrument on Unit Telescope 1 (UT1) of the VLT. A 1.0$\arcsec$ wide slit was used with a fixed position angle aligned on two field stars such that it would cover the position of \object{HLX-1} (Figure \ref{slit}). The \emph{Chandra} position of \object{HLX-1} has an offset of 0.14$\arcsec$
from the center of the slit, while the R-band counterpart detected by \citet{sor10} has an offset of 0.006$\arcsec$ from the center of the slit. As the error on both the \emph{Chandra} and R-band positions are 0.3$\arcsec$, both positions are well within the slit. This slit angle cuts through the bulge of the galaxy, missing the brightest part of the nucleus. Seeing conditions during the observations (as measured from the acquisition data) were good on the December observing dates (0.7 -- 0.9$\arcsec$), but were considerably worse on the November observing date (1.1 -- 1.5$\arcsec$).

A log of the spectroscopic observations can be found in Table \ref{obs}. To maximize the wavelength range in the red part of the optical spectrum we used the 300I grism, with the OG590 filter used to suppress contamination by the second spectral order. This results in a nominal wavelength range $\sim$5860 \AA~to 1.2 $\mu$m, though the signal redwards of 10500 \AA~is too weak to be useful.

\begin{table}
\begin{center}
\caption{VLT FORS2 Observation Log.\label{obs}}
\begin{tabular}{cc}
\tableline\tableline
Start Date & Exposure Time\\
(UT) & (s)\\
\tableline
26/11/2009 02:57 & 4 $\times$ 600 \\
05/12/2009 01:11 & 4 $\times$ 600 \\
06/12/2009 00:51 & 4 $\times$ 600 \\
07/12/2009 01:00 & 4 $\times$ 600 \\
\tableline
\end{tabular}
\end{center}
\end{table}

\section{Data Reduction \& Analysis}

The data were reduced by using the standard procedures in IRAF\footnote{http://www.iraf.net/} within the {\it ccdred} and {\em twodspec} packages, using bias frames and flat-fields taken the same night. The individual reduced 600 s spectra from the December observations were co-added, with the cosmic rays removed in the process. The data from 26 November 2009 were excluded from the summed spectrum, as the poor seeing on that night meant that the inclusion of those data did not increase the overall signal-to-noise of the data set.

\begin{figure}
\epsscale{0.65}
\plotone{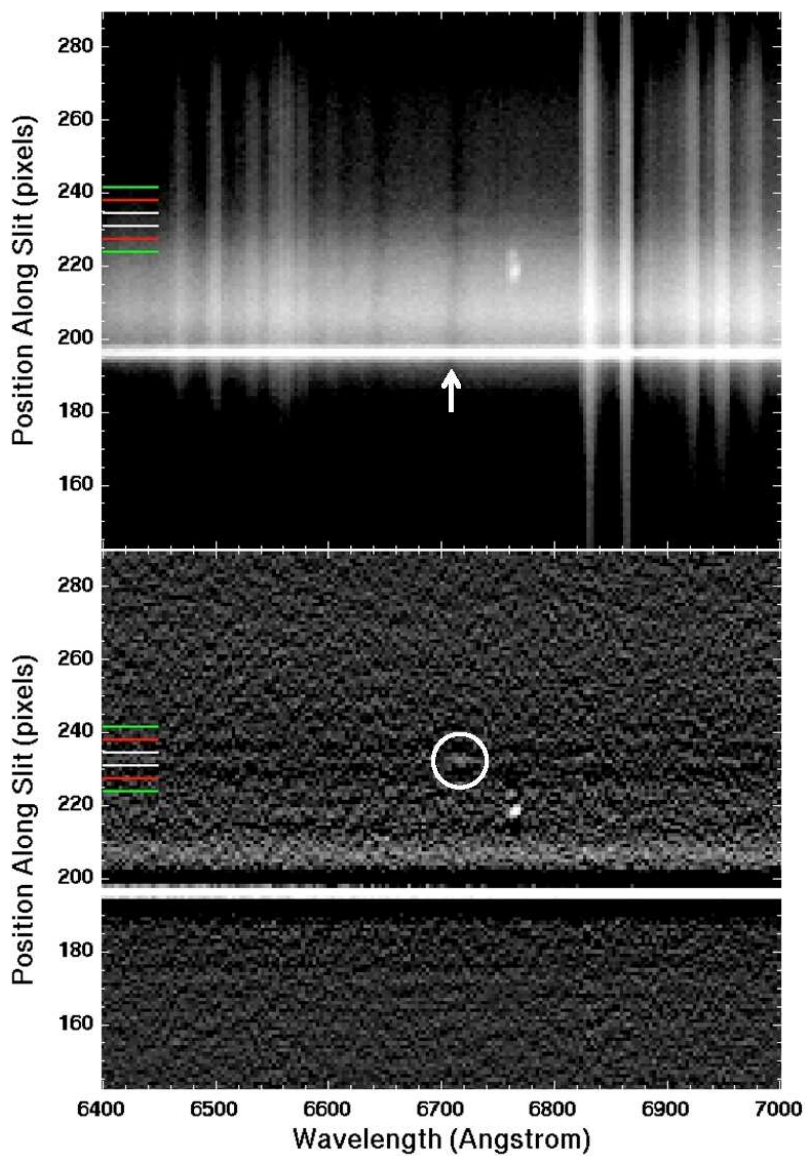}
\caption{\emph{Top:} the VLT 2-D spectrum of the HLX-1 field, with the ESO 243-49 H$\alpha$ absorption line indicated by the white arrow. \emph{Bottom:} 2-D spectrum after the diffuse galaxy emission has been removed via subtraction of a smoothed image. The circle indicates the position of the H$\alpha$ emission line in the HLX-1 spectrum. The nearby emission line source ($\sim$3$\arcsec$ away from HLX-1, so well outside the \emph{Chandra} error circle) appears to be a background galaxy. The sub-aperture covering the position of HLX-1 lies between the white horizontal lines to the left of the plots. The neighboring sub-apertures used to interpolate the background spectrum lie between the red and green and red and white lines.\label{2D}}
\end{figure}

We extracted the spectra from the merged 2-D spectrum (Figure \ref{2D}) by using the relatively bright continuum of the bulge to fit the shape of the trace function, and extracted 66 adjoining, equally sized sub-apertures along this trace. The pixel scale in the spatial direction of Figure \ref{2D} is 0.25\arcsec~pixel$^{-1}$. The sub-apertures were 0.9$\arcsec$ wide, giving a width slightly larger than the seeing full width at half maximum (FWHM) of the combined frame. At z = 0.0224 this corresponds to a physical scale of $\sim$0.4 kpc per sub-aperture, assuming a cosmology H$_0$ = 71 km s$^{-1}$ Mpc$^{-1}$, $\Omega_m$ = 0.27, $\Omega_\Lambda$ = 0.73. The series of sub-apertures chosen encompasses everything in between (and including) the two slit alignment stars. Furthermore, the position of the aperture and its subdivision into sub-apertures was chosen such that the expected position of the trace of \object{HLX-1}, determined through relative offsets with respect to the two slit alignment stars, was centred within one sub-aperture.  

In the following we will refer to the spectra extracted with these small apertures as sub-spectra. The sub-spectra were wavelength calibrated using He, HgCd and Ar arc lamp spectra. From a velocity cross-correlation of the wavelength calibrated arc spectrum with itself we measured a spectral resolution of 550 km s$^{-1}$ (FWHM). We boxcar smoothed the 2-D spectrum in the spatial direction with a 9 pixel box size, and subtracted this resulting image from the original 2-D spectrum to perform a crude subtraction of the bright galaxy profile. In this manner we subtracted the galaxy H$\alpha$ absorption line (likely stellar in origin), while not introducing artefacts due to the galaxy rotation curve (for which the scale is much larger than 9 pixels). 

\section{Results}

The galaxy bulge sub-spectra show strong absorption bands typical of a S0 type galaxy. The Na ID absorption feature can be seen in the blue end of the spectrum, giving a redshift for the galaxy of z = 0.0223 consistent with previous measurements \citep{afo05}. The H$\alpha$ line in absorption is also clearly detected at the same redshift, with an obvious curve in the 2-D spectrum due to the rotation of the galaxy spiral arms (Figure \ref{2D}). The absorption line visible at 6643 \AA~is the rest-frame 6497 \AA~blend (BaII, FeI and CaI) well known in late-G and early-K stars, which probably dominates in this galaxy and is consistent with the $\sim$5 Gyr old stellar population derived by \citet{sor10}. In addition to these features, a faint trace can be seen in the 2-D spectrum at the position of \object{HLX-1}, despite its proximity to the bright galaxy nucleus. The H$\alpha$ absorption feature is also present in the \object{HLX-1} trace, with visual inspection finding this feature to be partially filled in.

\begin{figure}
\epsscale{1}
\plotone{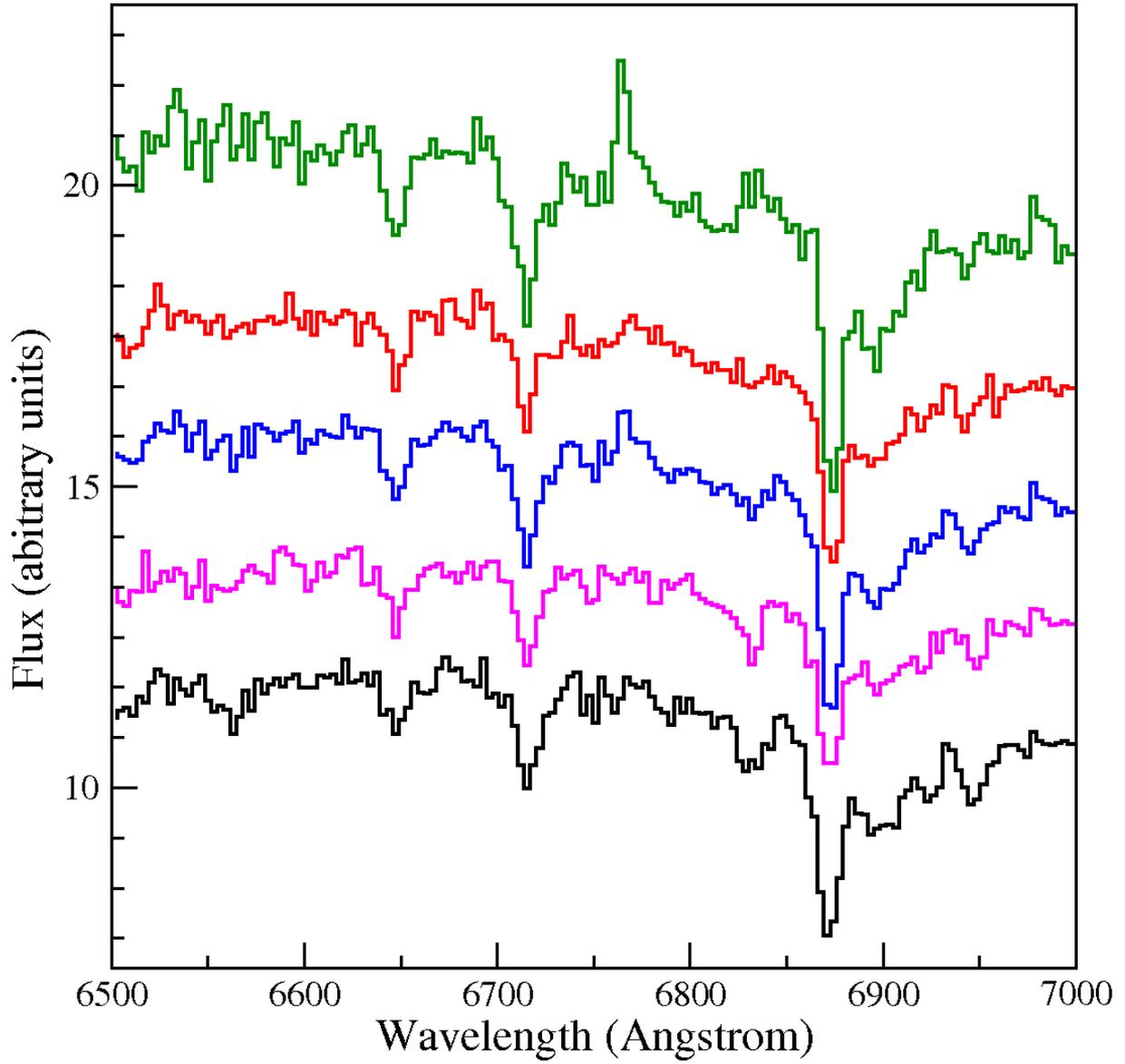}
\caption{Spectra from the neighboring sub-apertures (green, red, pink and black) that were used to derive the interpolated background spectrum at the position of HLX-1 (blue). A slight offset has been included to the arbitrary fluxes for the purpose of clarity.\label{bkgspec}}
\end{figure}

\begin{figure}
\epsscale{1}
\plotone{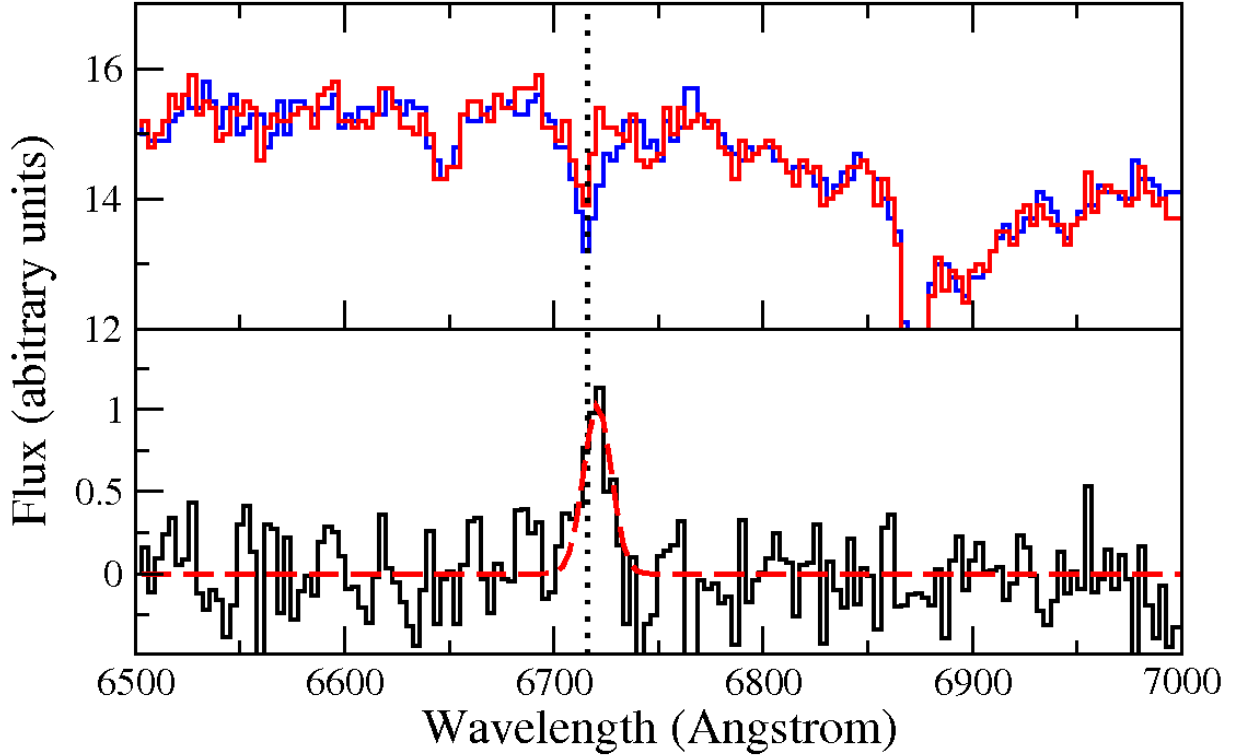}
\caption{ \emph{Top:} 1-D spectrum extracted from the position of HLX-1 (red) and the background spectrum (blue) interpolated from neighbouring sub-apertures. The absorption line at 6716 \AA~ is H$\alpha$ at the redshift of ESO 243-49. The signal-to-noise ratio around this absorption feature is $\sim$40 pixel$^{-1}$ in the HLX-1 sub-spectrum. \emph{Bottom:} background subtracted spectrum of HLX-1, showing line emission at 6721 \AA~with a significance of 11.3$\sigma$. The red dashed line indicates the Gaussian model fitted to the emission line. The dotted vertical line indicates the wavelength of the H$\alpha$ absorption line in the ESO 243-49 spectrum. In both plots the y-axis indicates arbitrary flux units. \label{subspectrum}}
\end{figure}

To confirm the association, we extracted sub-spectra from the position of \object{HLX-1} and from four sub-apertures taken from neighboring regions either side of the \object{HLX-1} position in the slit.  The fluxes of the neighboring sub-spectra were interpolated in count space as a function of wavelength to the \object{HLX-1} sub-spectrum using a spline function in order to estimate the background at the source position (Figure \ref{bkgspec}), and then subtracted from the source sub-spectrum. The resulting background subtracted spectrum reveals an emission line with an approximately Gaussian profile superimposed on a very weak continuum (Figure \ref{subspectrum}). The width of the spectral bins in both Figures \ref{bkgspec} and \ref{subspectrum} is 3.2 \AA. There is no other evidence of H$\alpha$ in emission anywhere else across the galaxy, despite the fact that the slit runs across part of the UV-bright region of ESO 243-49 in the low-resolution \emph{GALEX} and \emph{Swift} UVOT images \citep[see][]{web10}. This could indicate that star formation is uniform in this part of the galaxy, so that any H$\alpha$ signatures are subtracted out in the residual image, or that the emission features are too faint. However, without knowing the structure and scale of the UV emission we cannot be certain that the UV emission region falls in the slit. Higher resolution UV imaging with the \emph{Hubble Space Telescope} in cycle 18 has recently been awarded us,  so we should be able to answer this question in the near future.

To quantify the line significance, flux and wavelength, we used the sub-spectra. We performed a subtraction of the bright galaxy light from the \object{HLX-1} sub-spectrum by using the interpolated background spectrum described above. Flux calibration was performed on the resulting galaxy subtracted \object{HLX-1} spectrum using observations of the white dwarf standard star BPM 16274. We note that there are likely fairly large systematic errors associated with the procedure above: the bright galaxy brightness profile, though fairly smooth, changes on spatial scales similar to the sub-spectrum aperture sizes, so weighted linear interpolation of the adjacent sub-apertures can easily over or underestimates the contribution of the galaxy to the \object{HLX-1} sub-spectrum. These systematic errors make it very difficult to estimate the absolute flux of the continuum spectrum; however, the shape of the spectrum is not dependent on which sub-spectra were chosen for the background subtraction, or the interpolation method that was employed (e.g. spline vs linear interpolation functions). Hence, the presence of a significant emission line is not in doubt.

By fitting a Gaussian function we measure the wavelength of the emission line to be 6721.0 $\pm$ 1.1 \AA~and its FWHM to be 15.2 $\pm$ 2.3 \AA. This value is consistent with the instrumental resolution, and hence the line width should be taken as an upper-limit. The line significance was calculated from the 2-D spectrum once the smoothed spectrum was subtracted. We integrated the number of counts (ADUs) in the line in a rectangular aperture centered on the centroid of the Gaussian fit with a size twice the line FWHM in the wavelength direction and twice the seeing FWHM in the spatial direction. The error was calculated from the error in the resulting integrated counts and the standard deviation of the local background (i.e. the residuals from the subtraction). Thus the integrated significance of the H$\alpha$ line, taking into account the error introduced by the subtraction procedure, was calculated to be 11.3$\sigma$. The flux of the emission line is estimated to be 1.6 $\times$ 10$^{-17}$ erg s$^{-1}$ cm$^{-2}$. However, as the galaxy subtraction procedure introduces large systematic uncertainties, we adopt a conservative order of magnitude estimate for the line flux of $\sim$10$^{-17}$ erg s$^{-1}$ cm$^{-2}$. The flux of the continuum cannot be reliably measured with this data, as the low count rate is smaller than the errors introduced by interpolating the background.

\section{Discussion \& Conclusions}

The wavelength of the emission line detected in the \object{HLX-1} spectrum is consistent with H$\alpha$ at the redshift of ESO 243-49, with a velocity offset of $\sim$170 km s$^{-1}$ (considerably smaller than the galaxy rotation curve). The only other plausible identification of this line is [OII] for a source at z = 0.80; however, the F$_x$/F$_{opt}$ ratio of $\sim$1000 argues strongly against a background AGN. We therefore conclude that the line is most likely H$\alpha$ emission from an object gravitationally bound to ESO 243-49. 

A method for estimating the mass of black holes in low-redshift AGN has been established solely using observations of the H$\alpha$ emission line \citep{gre05}. A self-consistent picture could thus be developed in which the line width-luminosity-mass relation seen for AGN can be extrapolated down to lower black hole masses. At the galaxy redshift the line luminosity is $\sim$2 $\times$ 10$^{37}$ erg s$^{-1}$, so if we assume that the H$\alpha$ scaling relationship holds for lighter black holes, the line luminosity and FWHM imply a black hole mass of $\sim$1500 M$_\odot$ for \object{HLX-1}. However, as the line is unresolved in our spectrum, this mass estimate should be taken as an upper-limit.  In addition, this relationship would only hold if the origin of the H$\alpha$ emission is the same as in AGN (i.e. the broad line region).  Alternatively, the H$\alpha$ line could be produced by the environment in which HLX-1 lies, perhaps a star cluster or photo-ionized/shock-ionized gas nebula, in which case this method would not apply and the mass upper-limit would not stand. The large uncertainties inherent in the process of subtracting the diffuse emission from the galaxy mean that we are not able to set any meaningful limits on the continuum flux. Thus, with the data at hand we cannot place any new constraints on the nature of the optical counterpart (e.g. star cluster, globular cluster, nucleated dwarf galaxy etc.), beyond what can be done using the photometry already reported in \citet{sor10}.

The detection of a H$\alpha$ emission line at a redshift of z = 0.0223 definitively places \object{HLX-1} within ESO 243-49, confirming the 10$^{42}$ erg s$^{-1}$ luminosity. As such \object{HLX-1} continues to provide the strongest case to date for the existence of intermediate mass black holes in at least some ULXs. The large-scale temporal and spectral variability of this object therefore provides us with a unique opportunity to study an intermediate mass black hole in different accretion regimes, allowing us to make direct comparisons with the populations of stellar mass and super-massive black holes. An ongoing monitoring program in X-rays with the \emph{Swift} observatory and planned follow-up observations at other wavelengths should continue to shed light on the nature and environment of this enigmatic object.

\acknowledgments

We thank M. C. Miller and the anonymous referees for their comments that helped to improve this paper. We thank Tim de Zeeuw for according us the VLT DDT observations. Based on observations made with ESO Telescopes at the Paranal Observatory under programme ID 284.D-5008. IRAF is distributed by the National Optical Astronomy Observatories, which are operated by the Association of Universities for Research in Astronomy, Inc., under 
cooperative agreement with the National Science Foundation. S.A.F., K.W. and T.J.M. acknowledge STFC funding. T.J.M. thanks the European Union FP7 for support through grant 215212 ÒBlack Hole UniverseÓ. M.S. is supported in part by Chandra grants AR9-0013X and GO9-0102X. 



{\it Facilities:} \facility{VLT (FORS2)}.



\clearpage



\end{document}